\documentclass[]{jfm}
\usepackage{graphicx}
\usepackage{epstopdf, epsfig}
\usepackage{multirow}
\usepackage{mathrsfs,amsmath}
\usepackage{bm}
\usepackage{amssymb}
\usepackage{multirow,bigdelim}
\usepackage{upgreek}
\usepackage{units}
\usepackage{array,multirow}
\usepackage{booktabs}
\usepackage{psfrag}
\usepackage{tikz}
\usetikzlibrary{arrows.meta}
\usepackage{color}
\usepackage{xcolor}
\usepackage{threeparttable}	
\usepackage{dcolumn}		
\newcolumntype{d}[1]{D{.}{.}{#1}}
\usetikzlibrary{decorations}
\usetikzlibrary{decorations.markings}
\usetikzlibrary{calc,decorations.pathreplacing}
\usepackage{lipsum}
\usepackage{mathtools}
\usepackage{xfrac}
\usepackage{ar}
\usepackage{enumitem}
\usepackage{url}
\usepackage{subcaption}
\usepackage{mwe}
\usepackage{float}
\usepackage{lmodern}
\usepackage{soul}

\shorttitle{Scale-dependent inclination angle of turbulence in stratified ASL's}
\shortauthor{Xuebo Li et al.}

\title{\textcolor{black}{Scale-dependent inclination angle of turbulent structures in stratified atmospheric surface layers}}

\author{Xuebo Li\aff{1,2}, Nicholas Hutchins\aff{2}, Xiaojing Zheng\aff{1}\corresp{\email{xjzheng@lzu.edu.cn}}, \\ Ivan Marusic\aff{2} \and Woutijn J. Baars\aff{3}}

\affiliation{\aff{1}Center for Particle-Laden Turbulence, Department of Mechanics, Lanzhou University, Lanzhou 730000, People’s Republic of China
\aff{2}Department of Mechanical Engineering, University of Melbourne, VIC 3010, Australia
\aff{3}Faculty of Aerospace Engineering, Delft University of Technology, 2629 HS, The Netherlands}
\begin{document}

\maketitle
\begin{abstract}
A large-scale spanwise and wall-normal array of sonic anemometers in the atmospheric surface layer is used to acquire all three components of instantaneous fluctuating velocity as well as temperature in a range of stability conditions. These data permit investigation of the three-dimensional statistical structure of turbulence structures. The present work extends the view of a self-similar range of wall-attached turbulence structures to the atmospheric surface layer under unstable and near-neutral stability conditions, and includes the statistical structure in both the wall-normal and spanwise directions in relation to the streamwise wavelength. Results suggest that the self-similar wall-attached structures have similar aspect ratios between streamwise/wall-normal scales and streamwise/spanwise scales such that $\lambda_x/\Delta z : \lambda_x/\Delta y \approx 1$ for both near-neutral and unstable conditions. By analysing the phase shift between synchronized measurements, in the spectral domain, it is quantified how the structure inclination angle varies with stability. Under the most unstable conditions, coherent structures of $\lambda_x/\delta = 1$ are inclined at angles as high as $65^\circ$ relative to the solid boundary, while larger scales of $\lambda_x/\delta = 6$ exhibit inclination angles of approximately $35^\circ$. For near-neutral stability conditions, the angle tends towards $12^\circ$ for all scales. It is noted that in the near-neutral condition, the structure inclination angle and the aspect ratio---and thus the statistical modeling of coherent structures in the ASL---are highly sensitive to the value of the stability parameter.
\end{abstract}

\section{Introduction}\label{sec:intro}
\begin{figure} 
\vspace{0pt}
\centering
\includegraphics[width = 0.999\textwidth]{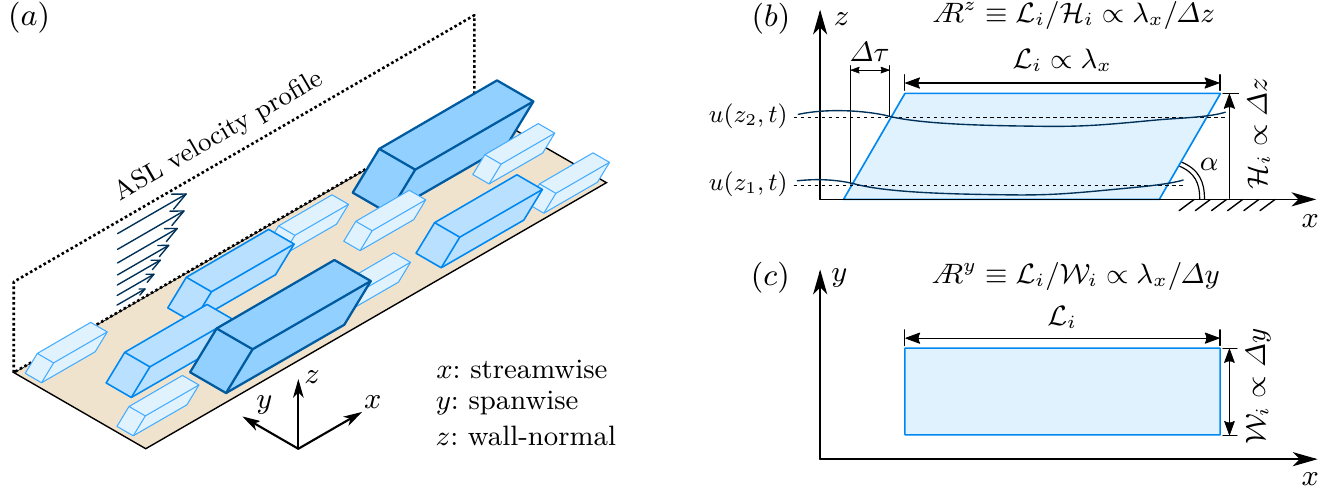}
\caption{($a$) Isometric view of three hierarchies of self-similar wall-attached eddies (simplified as slanted cuboids) in the logarithmic region of an ASL. ($b$) $x$,$z$--plane and ($c$) $x$,$y$--plane of one structure. Here, $\mathcal{L}_i$, $\mathcal{W}_i$ and $\mathcal{H}_i$ denote the streamwise, spanwise and wall-normal extent of the $i^{\rm th}$ hierarchy structure.}
\label{intro_sketch}
\end{figure}

\citet{townsend1976structure} proposed a conceptual model for wall-bounded turbulence, the attached eddy hypothesis (AEH), which idealizes structures as a collection of inertia-driven self-similar eddies that are randomly distributed in the plane of the wall. Details of key assumptions and limitations associated with the AEH are covered in a recent review by \citet{marusic2019attached}. Based on the AEH, \citet{perry1982mechanism} proposed that coherent wall-attached eddies scale with the distance from the wall $z$, and their heights comprise a geometrical progression. Evidence in support of a self-similarity and wall-scaling of wall-attached vortices has been reported in recent turbulent boundary layer (TBL) studies \citep[\emph{e.g.},][]{jimenez2012cascades, hwang2015statistical, baars2017self}. Figure~\ref{intro_sketch} shows an idealization of a self-similar hierarchy of wall-attached structures within the logarithmic region of a TBL \citep[][]{baidya2019simultaneous,marusic2019attached,deshpande2019streamwise}. Here, we consider three hierarchy levels of randomly positioned regions of coherent velocity fluctuations with each hierarchy shown in a different color. For simplicity, we consider the volume of influence of eddies, in each level, to be characterized by $\mathcal{L}_{i}$, $\mathcal{W}_{i}$ and $\mathcal{H}_{i}$ in the $x$, $y$ and $z$ directions, respectively, with $i=1,2,3$ denoting the $i^{\rm th}$ hierarchy level. Figures~\ref{intro_sketch}($b$) and~\ref{intro_sketch}($c$) denote the aspect ratios in the streamwise/wall-normal plane \AR$^{z}\equiv \mathcal{L}_i/\mathcal{H}_i\propto \lambda_{x}/\Delta{z}$ and in the streamwise-spanwise plane \AR$^{y}\equiv \mathcal{L}_i/\mathcal{W}_i\propto \lambda_{x}/\Delta{y}$, respectively. \citet{baars2017self} reported that in the neutral laboratory zero-pressure gradient TBL the self-similarity is described by a streamwise/wall-normal aspect ratio of $\lambda_{x}/\Delta{z} \approx 14$. A recent study by \citet{baidya2019simultaneous}, in high Reynolds number pipe and boundary layer flows, indicated that the self-similar wall-attached structures follow a three-dimensional aspect ratio of 14:1:1 in the streamwise, spanwise and wall-normal directions, respectively. More recently, \citet{krug2019vertical} explored the coherence for both velocity and temperature signals in the ASL. They found that the streamwise/wall-normal aspect ratio (\AR$^{z} \equiv \lambda_x/\Delta z$) decays with a logarithmic trend with increasing unstable thermal stratification; spanwise information was not explored in their study.

Coherent structures reported in TBLs have been associated with characteristic inclination angles because of the mean shear. In the idealized (statistical) view of figure~\ref{intro_sketch}($b$), inclination angle $\alpha$ in the $x$,$z$--plane reflects a phase shift $\Delta \tau$ in time series of velocity fluctuations at different $z$. \citet{perry1982mechanism} used a vortex skeleton approach and the Biot-Savart law to determine the inviscid velocity field of a representative eddy, termed the $\Lambda$-vortex. In neutral TBLs, hairpin vortices have commonly been invoked as the representative eddy, and \citet{adrian2000vortex} suggested that these vortices are arranged together in groups called vortex packets \citep[for a comprehensive review of hairpin structures and their generating mechanisms, see][]{adrian2007hairpin}. \citet{christensen2001statistical} found that the sequence of individual vortex heads forms an interface or shear layer that is, statistically, inclined away from the wall at angles between $12^\circ$ and $20^\circ$. Laboratory results indicate the most probable inclination angle to be around $10^\circ$ to $15^\circ$ \citep[][]{christensen2001statistical,adrian2000vortex,baars2016spectral}. In the neutral surface layer, inclination angles ranging from 10$^\circ$ to 20$^\circ$ have been reported \citep[][]{boppe1999large, carper2004role,chauhan2013structure,liu2017variation}. Also, \citet{marusic2007reynolds} demonstrated the invariance of the inclination angle in wall-bounded flows with zero buoyancy (neutral conditions) over a wide range of Reynolds number through laboratory and field experiments.

The aforementioned studies all refer to the near-neutral case. However, in studies of the atmospheric surface layer (ASL) it has been observed that the inclination angle changes drastically under different stability conditions \citep[with steeper angles in increasingly buoyant cases, see][]{chauhan2013structure, liu2017variation, lotfy2018effect}. The thermal stability of the ASL is generally characterized by the Monin-Obukhov stability parameter $z_s/L$ \citep[][]{obukhov1946turbulence,monin1954basic}, where $L=-u_{\tau}^3{\overline{\theta}}/\kappa{\overline{w\theta}}{g}$ is the Obukhov length, $\kappa=0.41$ the von K\'{a}rm\'{a}n constant, $g$ the gravitational acceleration, $\overline{w\theta}$ the surface heat flux with $w$ and $\theta$ the fluctuating wall-normal velocity and temperature components, $\overline{\theta}$ the mean temperature, $u_{\tau}$ the friction velocity, and $z_s$ the reference height for evaluating this parameter. \citet{chauhan2013structure} found that under stable conditions, the inclination angle of structures reduced below the near-neutral angle to approximately $10^\circ$. Recently, \citet{salesky2020coherent} introduced an additional parameter to account for the loading and unloading of surface layer flux-gradient relations imposed by the passage of large-scale motions (LSMs).  Meanwhile, \citet{salesky2020revisiting} developed a prognostic model for large-scale structures, where the inclination angle is the sum of the inclination angle observed in a neutrally stratified wall-bounded turbulent flow and the stability-dependent inclination angle of the wedge. \citet{baars2016spectral} indicates that in the neutral case, and for all scales $\lambda_x/\delta > 0.5$, the coherent scales obey a virtually constant inclination angle. In unstable conditions in the atmosphere, positive buoyancy lifts the structure away from the surface leading to an increase in the statistical inclination angle \citep[as averaged across all scales, see][]{chauhan2013structure,liu2017variation}. Now, in the unstable case, the dominance of buoyancy over shear is a function of wall-normal height, and hence one expects the inclination angle to be scale-dependent.

Since the coherent structure in the ASL has a strong relationship with the stability parameter, this paper will specifically address the influence of stability on:  (1) the streamwise/wall-normal aspect ratio \AR$^{z}\equiv\lambda_{x}/\Delta z$ and the streamwise/spanwise aspect ratio \AR$^{y}\equiv\lambda_{x}/\Delta y$ in \S\,\ref{sec:ratio}, and (2) the scale-dependent angle $\alpha$ in \S\,\ref{sec:angle}, particularly under unstable conditions. Statistical relations for the aspect ratio and inclination angle for coherent turbulence fluctuations in the ASL are particularly relevant when analysing wind loading in the field of wind engineering \citep[see][]{davenport:1961phd,davenport:2002a}.

\section{Turbulence dataset of the atmospheric surface layer}\label{sec:data}
\subsection{QLOA facility and available data}\label{sec:qloa}
\begin{figure} 
\vspace{0pt}
\centering
\includegraphics[width = 0.999\textwidth]{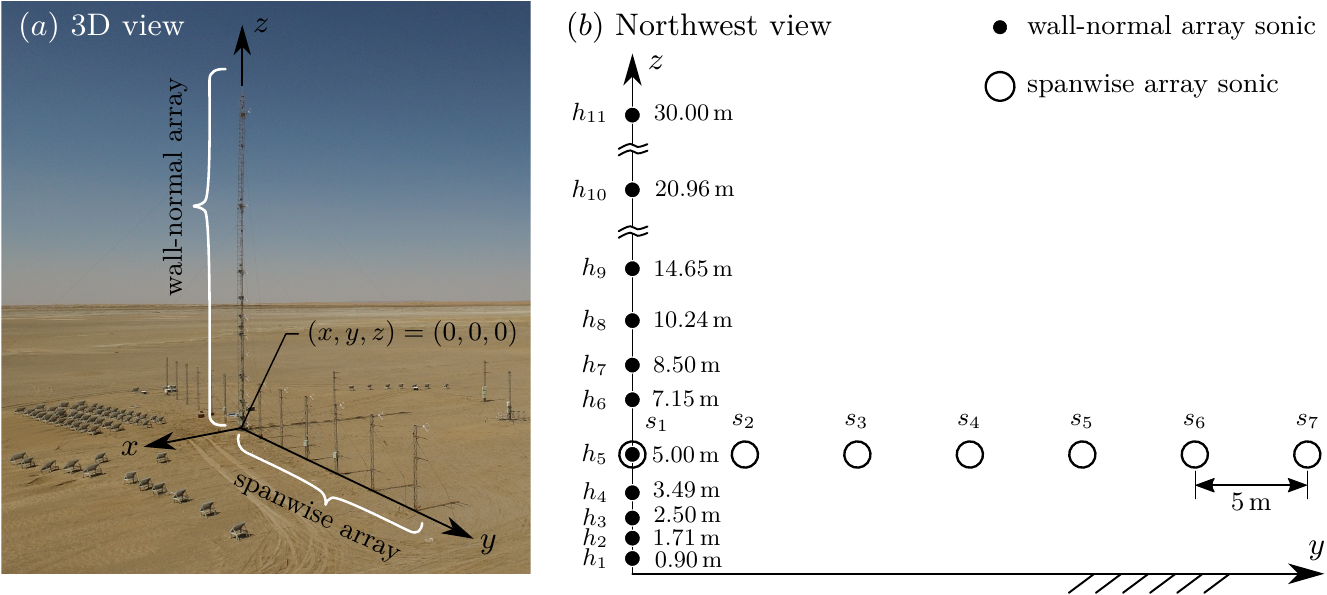}
\caption{($a$) Three-dimensional view of the measurement setup at the QLOA site. ($b$) Northwest view of the sonic anemometer array; Campbell CSAT3B and Gill Instruments R3-50 sonics were installed at positions $h_1$ to $h_{11}$ and $s_2$ to $s_7$, respectively.}
\label{experiment_site}
\end{figure}

The measurement data used throughout this article were acquired at the QLOA facility in western China, Gansu province during three-month long measurement campaigns over two years (March to May in 2014 and 2015). The QLOA consists of wall-normal and spanwise arrays of sonic anemometers, performing synchronous measurements of the three-dimensional turbulent flow field. Sonic anemometers (Gill Instruments R3-50 installed from $s_2$ to $s_7$ and Campbell CSAT3B installed from $h_1$ to $h_{11}$, figure~\ref{experiment_site}) were employed to acquire the three components of velocity, as well as the static temperature, at a sampling frequency of 50\,Hz. Continuous observations were conducted at the QLOA site for a duration of more than 3000 hours, from which 89 hours of data were selected to analyze the characteristics of the large-scale coherent structures under different stratification stability conditions. The wall-normal array consists of 11 sonic anemometers that were placed with a logarithmic spacing on a vertical radio-type tower. The spanwise array covered an overall distance of 30\,m with 7 anemometers that were placed at a constant height of $z = 5$\,m, with an equi-distant spanwise spacing of 5\,m. The spanwise and wall-normal coordinates for each of the 17 anemometers are provided in figure~\ref{experiment_site}($b$). It should be noted that the first sonic anemometer in the spanwise array ($s_1$) also functions as the fifth on the main tower ($h_5$), which means we have 7 available anemometers in the spanwise array. The friction velocity $u_{\tau}$ is inferred from $u_{\tau}=(-\overline{uw})^{1/2}$ at $z = 5$\,m (calculated by the mean value from 7 sonic anemometers in the spanwise array). We assume an estimate for the surface-layer thickness of $\delta = 60$\,m, following \citet{hutchins2012towards}. The 89 hours of data remaining after preselection include 69 hours of unstable data ($z_s/L<-0.01$), 10 hours of near neutral data ($-0.01\leq{z}_s/L<0.01$) and 10 hours of stable data ($z_s/L>0.01$). Recall that $z_s$ is the reference height used to define the stability parameter $z_{s}/L$. For the benefit of comparison with previous works \citep{chauhan2013structure,liu2017variation,krug2019vertical}, the majority of our work uses $z_s = 2.5$\,m, unless otherwise specified. The demarcation of $z_s/L = 0.01$ to distinguish between neutral and unstable thermal stratification is commonly found in the literature, but for all analysis in this paper we present results as a function of $z_s/L$. The preselection criteria included: wind direction (the wind direction had to be aligned with the $x$ axis of the anemometer to within $\pm 30^\circ$) and steadiness \citep[statistically steady conditions based on the high-quality requirement by][]{fiken2004post}. A de-trending operation is also added (to remove the large-scale synoptic trend). See \citet{hutchins2012towards} and \citet{wang2016very} for full details of the preselection criteria. 

\subsection{Processing method}\label{sec:processing}
\begin{figure} 
\vspace{0pt}
\centering
\includegraphics[width = 0.999\textwidth]{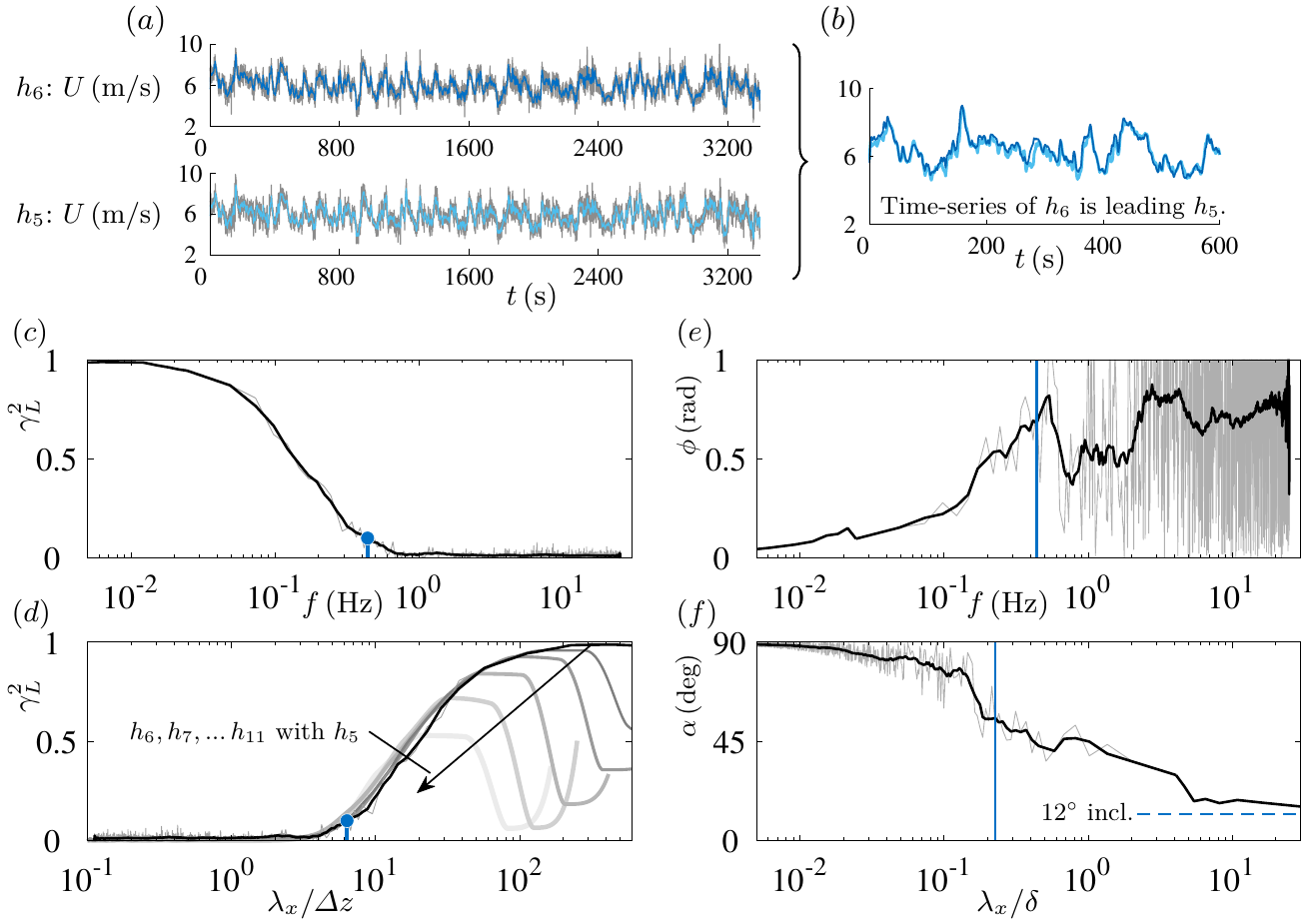}
\caption{($a$) An example of time series data for $z_s/L = -0.52$ at heights $h_5$ and $h_6$. The blue lines are low-pass filtered at $f = U/\delta$. ($b$) Portion of the filtered signals for inspecting the coherence and time shift. ($c$) Coherence spectrum for height $h_6$, relative to $h_5$, as function of temporal frequency. ($d$) Coherence spectrum for all heights $h_6$\,...\,$h_{11}$, again relative to $h_5$. Here the abscissa is converted to a spatial wavelength with the mean velocity at height $z$. ($e$) Scale-dependent phase of the cross-spectrum between $h_5$ and $h_6$, which is converted to a physical inclination angle $\alpha$ as a function of a spatial wavelength in sub-figure ($f$). The blue dots and blue solid lines in ($c,f$) indicate the frequency or wavelength corresponding to $\gamma _L^2=0.1$. Filtered spectra (black line), overlaid on the grey raw spectra, utilize a bandwidth moving filter of 25\%.}
\label{fig:processing}
\end{figure}

Two-point correlations can be computed on a per-scale basis in the Fourier domain using the linear coherence spectrum (LCS). For the fluctuating streamwise velocity signals $u$ the LCS is defined as:
\begin{equation}	
    \gamma_{L}^{2}(z,z_{\rm lcs};\lambda_{x})\equiv\frac{|\langle{\widetilde{U}(z;\lambda_{x})\widetilde{U}^{*}(z_{\rm lcs};\lambda_{x})}\rangle|^{2}}{\langle{|\widetilde{U}(z;\lambda_{x})|^{2}}\rangle\langle{|\widetilde{U}(z_{\rm lcs};\lambda_{x})|^{2}}\rangle}=\frac{|\phi_{uu}^{'}(z,z_{\rm lcs};\lambda_{x})|^{2}}{\phi_{uu}(z;\lambda_{x})\phi_{uu}(z_{\rm lcs};\lambda_{x})}.
\label{equ:spectrum}
\end{equation}
Here, $\widetilde{U}(z;\lambda_{x})=\mathcal{F}[u(z)]$ is the Fourier transform of $u(z)$, in either $x$ or time. The spatio-temporal transformation uses Taylor's hypothesis \citep{taylor1938spectrum}, where the local mean velocity is taken as the convection velocity. The asterisk $*$ indicates the complex conjugate, $\langle\rangle$ denotes ensemble averaging and $||$ designates the modulus. Scale-dependent phase information is explicitly embedded in the phase of the cross-spectrum $\phi_{uu}^{'}$. For (\ref{equ:spectrum}), the LCS is defined based on $u$ at two positions ($z_{\rm lcs}$ and $z$), separated in the wall normal direction by $\Delta z \equiv z - z_{\rm lcs}$. This coherence can also be computed across all other measured signals ($v$ and $\theta$) and also across spanwise separations $\Delta y$. The reference signal's height is denoted with the subscript `lcs' and is thus stated as $z_{\rm lcs}$. Since the LCS considers the magnitude of the complex-valued cross-spectrum, only the magnitude of coherence is considered (phase is covered later). Based on assumptions from the AEH \citep[][]{baars2017self,krug2019vertical}, the coherence magnitude within a self-similar region in $\lambda_x,z$--space is expected to adhere to
\begin{equation}
    \gamma_{L}^{2}=C_{1}{\ln}\left( \frac{\lambda_{x}}{\Delta z} \right) +C_{2},
\label{EQgamma}
\end{equation}
\noindent from which the statistical aspect ratio (in this case streamwise/wall-normal) then follows
\begin{equation}
    \AR^z = \left. \frac{\lambda_x}{\Delta z}\right\vert_{\gamma_{L}^2=0}=\exp\left(\frac{-C_{2}}{C_{1}} \right).
\label{EQgamma2}
\end{equation}
Here, $C_1$ and $C_2$ are fitted parameters. Figure \ref{fig:processing} shows an example of data obtained from the ASL to illustrate the process of the coherence spectrum. Figure \ref{fig:processing}($a$) indicates the raw data for the streamwise velocity collected at $h_5 = 5$\,m and $h_6 = 7.15$\,m, under unstable conditions with  $z_{s}/L=-0.52$. A shorter time-history of corresponding filtered signals are shown in figure~\ref{fig:processing}($b$), evidencing that signal $h_6$ leads $h_5$. Thus, a coherent velocity fluctuation is first sensed at the higher wall-normal location as a result of the structure inclination angle. The LCS for $h_5$ and $h_6$ is presented in figure~\ref{fig:processing}($c$) as a function of temporal frequency, as computed from the time series data. Using Taylor’s frozen turbulence hypothesis, the frequency axis can be converted to a streamwise wavelength: $\lambda_x \equiv U_c/f$. Here, $U_{c}$ is a convective speed, taken as the mean velocity at local height $z$. The coherence spectrum of figure~\ref{fig:processing}($c$) can now be presented as a function of the streamwise wavelength, relative to the wall-normal separation distance $\Delta z$, as shown in figure~\ref{fig:processing}($d$). In addition to the coherence spectrum for $h_5$ and $h_6$, the coherence spectra for all heights above $h_6$ (relative to $h_5$ again) are also shown to illustrate the wall-similarity that is to be investigated. \citet{krug2019vertical} noticed that only data at neutral and unstable thermal stratification conditions complied with (\ref{EQgamma}), and thus an aspect ratio was only found for those conditions (a similar conclusion was reached for our QLOA site data and therefore we do not consider stable stratification with $z_s/L > 0$). In addition, \citet{krug2019vertical} found that the self-similar scaling applies also to fluctuations of the spanwise velocity $v$ and the static temperature $\theta$. \citet{baidya2019simultaneous} demonstrated that a scaling similar to (\ref{EQgamma}) and (\ref{EQgamma2}) occurs in the spanwise direction, resulting in a streamwise/spanwise aspect ratio \AR$^y$ for the self-similar structure.

Scale-dependent phase information is explicitly embedded in the phase of the cross-spectrum $\phi_{uu}^{'}$, given by
\begin{equation}
    \Phi\left(z,z_{\rm lcs};\lambda_x\right)= \tan^{-1}\left(\frac{\Imag [\phi_{uu}^{'}(z,z_{\rm lcs};\lambda_x) ]}{\Real\left[\phi_{uu}^{'}(z,z_{\rm lcs};\lambda_x)\right]}\right).
\label{equ:phase}
\end{equation}
The phase spectrum aids in assessing the temporal shift between signals. Phase $\Phi(f)$ in (\ref{equ:phase}) is shown in figure~\ref{fig:processing}($e$) and can be used to extract a scale-by-scale inclination angle $\alpha$ (as shown in figure~\ref{fig:processing}f). That is, the temporal shift $\tau=\Phi(f)/(2\pi{f})$ where $f$ is the mode frequency and aids in computing the physical inclination angle through $\alpha=\tan[\Delta{z}/(\tau U_{c})]$. For the spectral analysis, the highest frequency resolved is set by the Nyquist frequency $f_s/2 = 25$\,Hz, where $f_s = 50$\,Hz is the sampling frequency. The lowest frequency is dictated by the interval length $I$ used in the spectral analysis and the longest internal used was $I = 2^N$ samples with $N = 15$ (an interval length of $\approx 650$\,s). A composite approach with varying interval length ($N = 8\,...\,15$) was used to generate the full spectra with as many ensembles as possible for the higher frequency portions of the spectrum (for $N = 8$ a total of around 1000 ensembles were used).

\section{Results}\label{sec:results}
\subsection{Stability dependence of aspect ratio}\label{sec:ratio}
\begin{figure} 
\vspace{0pt}
\centering
\includegraphics[width = 0.999\textwidth]{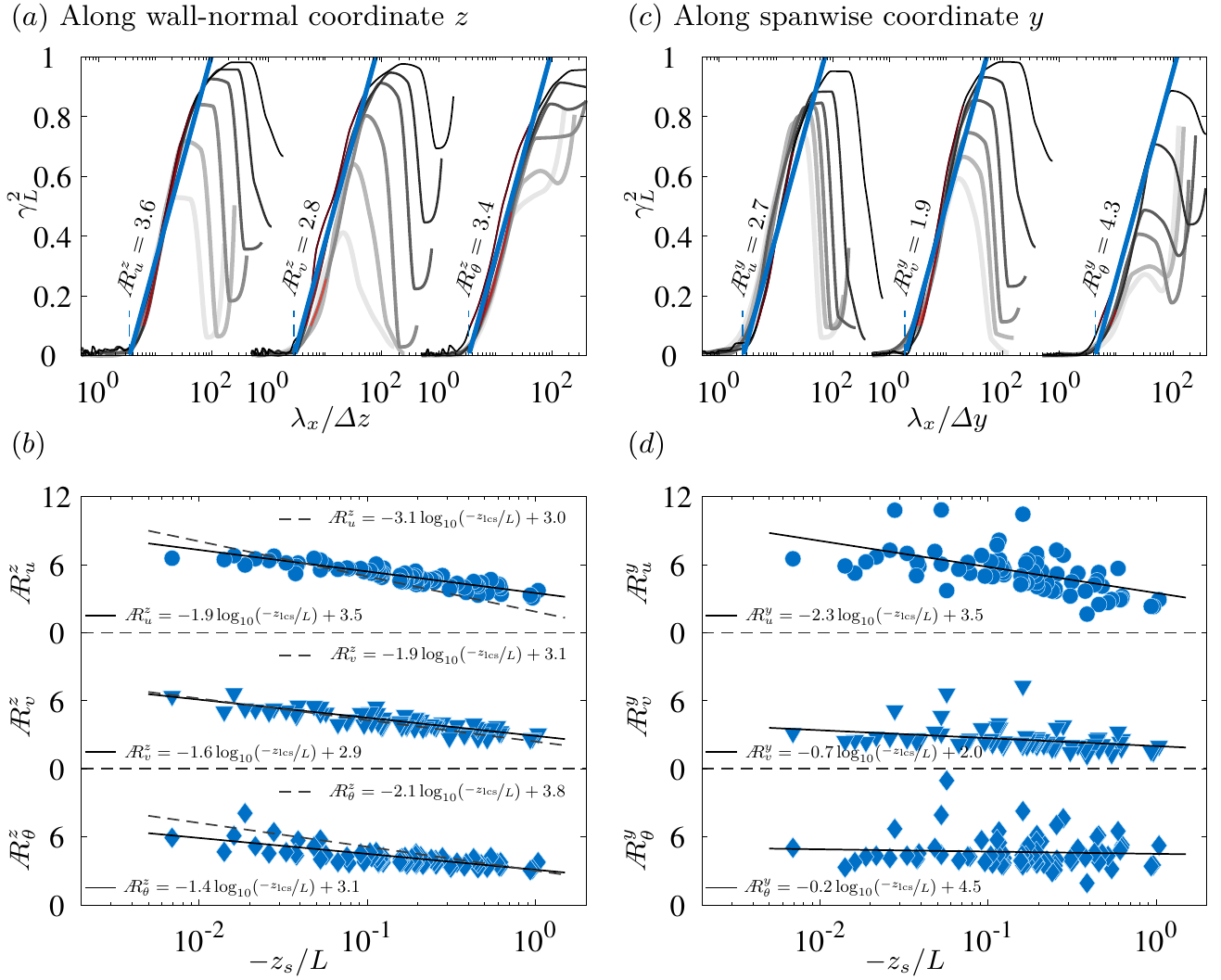}
\caption{$\gamma^2_L$ in the range $2.15\,{\rm m} \leq \Delta{z} \leq 25$\,m and $5\,{\rm m} \leq \Delta{y} \leq 30$\,m (with increasing $\Delta{z}$ and $\Delta{y}$ indicated by lighter shades of grey) for the ($a$) wall-normal coordinate $z$ and ($c$) spanwise coordinate $y$, respectively, for the unstable case with $z_s/L = -0.52$. Here, the range between $\Delta z_{5,6} = h_6 - h_5 = 2.15$\,m and $\Delta z_{5,11} = h_{11} - h_5 = 25$\,m; $\Delta y_{1,2} = s_2 - s_1 = 5$\,m and $\Delta y_{1,7} = s_7-s_1 = 30$\,m. The blue line is a fit to obtain the aspect ratio according to (\ref{EQgamma}) with $C_{1}=0.302$ fixed; the fitting region used is bounded by $\gamma_{L}^2>0.1$ and $\lambda_x < 100$\,m and is indicated in red lines. Subscript $i=u,v,\theta$ in \AR$_{i}^k$ signifies the aspect ratio for streamwise, spanwise velocity components as well as temperature component; superscript $k=z,y$ in \AR$_{i}^k$ indicates either the streamwise/wall-normal or streamwise/spanwise aspect ratio. ($b,d$) Streamwise/wall-normal and streamwise/spanwise aspect ratios, respectively, as a function of $z_s/L$. The blue dots are our ASL results and the black solid lines denote the semi-log fitting. The dashed lines in ($b$) come from \citet{krug2019vertical} with $z_{\rm s}=2.14$\,m and $z_{\rm lcs}=1.41$\,m. Note that in this work $z_{\rm lcs}=z_s=5$\,m and so the dashed lines of \citet{krug2019vertical} were shifted along the $z_s$ axis to compare the trends at matched $-z_s/L$.}
\label{result_gamma}
\end{figure}

The linear coherence spectrum for $u$, $v$ and $\theta$ as a function of $\lambda_x/\Delta z$  and $\lambda_x/\Delta y$ for the unstable case with $z_s/L = -0.52$ is given in figures~\ref{result_gamma}($a$) and \ref{result_gamma}($c$), respectively. As reported by \citet{krug2019vertical}, the LCS collapse on one common curve over a range of $\lambda_x/\Delta z$ and $\lambda_x/\Delta y$. By fitting (\ref{EQgamma}) to these regions to obtain $C_1$ and $C_2$, the aspect ratio $\AR$ can be assessed. For this particular unstable case, we might expect the positive buoyancy to cause the self-similar structures in the hierarchy to lift more aggressively from the wall, extending the wall-normal coherence for a given $\lambda_x$ scale, and hence reducing \AR$^z_u$. Indeed, for the unstable case considered in figure~\ref{result_gamma}($a$) this yields an $\AR_u^z$ that is significantly lower ($ \AR^z_{u}= 3.6$) than the value of \AR$^z_{u}\approx14$ reported for laboratory neutral conditions \citep{baars2017self} and in close agreement with the aspect ratio reported in \citet{krug2019vertical} for similar values of the stability parameter. The resulting aspect ratios for $u$, $v$ and $\theta$ from all 79 hour datasets (covering a range of stabilities from $0.007 \leq -z_{\rm lcs}/L \leq 1.04$ are plotted as a function of the stability parameter in figure~\ref{result_gamma}($b$) for streamwise/wall-normal aspect ratios $\AR^z$ and  in figure~\ref{result_gamma}($d$) for streamwise/spanwise aspect ratios $\AR^y$. In all cases, a clear trend emerges between aspect ratio and stability parameter, and a log-linear trend is fitted to the extracted data (black  solid curves). These fitted trends are consistent with those of \citet{krug2019vertical} (black dashed curves in figure~\ref{result_gamma}$b$) indicating that the self-similar scaling under near-neutral and unstable conditions is a universal phenomenon. As an extension to the results of \citet{krug2019vertical}, the streamwise/spanwise aspect ratios of figure~\ref{result_gamma}($d$) seem to also exhibit log-linear trends, although in these cases the scatter in results is greater.

\citet{baidya2019simultaneous} indicates that the aspect ratio \AR$^z_{u}:$\AR$^y_{u}=1:1$ in the laboratory neutral condition. The results shown in figure~\ref{result_gamma}($b,d$) are also supportive of this. The curve fits to the data suggest that $\AR_u^z:\AR_u^y = 0.91:1$ for the near-neutral case  ($\vert z_s/L \vert < 0.03$) changing slightly to 0.96:1 for the strong unstable  ($z_s/L < -0.40$) stability conditions. Moreover, the aspect ratio for $u$ shows that  $\lambda_x:\Delta y:\Delta z\approx 4.25:0.96:1 $ under the strong unstable condition, demonstrating that positive buoyancy has a lifting effect, increasing the size of coherent structures in the wall-normal and spanwise directions relative to its streamwise extent. The data here indicate that these self-similar eddies for $u$ follow an aspect ratio of $\lambda_x:\Delta y:\Delta z\approx 6.4:0.91:1 $ in the near-neutral condition ($\vert z_s/L \vert < 0.03$). It is worth highlighting again that the aspect ratio is sensitive to even very weakly unstable conditions. Therefore, the value for $u$ measured in the ASL is less than the result $\lambda_x/\Delta{z} = 14$ from \citet{baars2017self} and \citet{baidya2019simultaneous} in neutral laboratory conditions, as was also noticed by \citet{krug2019vertical}, whose prediction implies that $\lambda_x/\Delta z = 14$ will only be attained for $\vert z_s/L \vert \approx 0.0003$ (but for the QLOA dataset we only have data for $\vert z_s/L \vert \geq 0.007$).

\subsection{Stability dependence of structure inclination angle}\label{sec:angle}
\begin{figure} 
\vspace{0pt}
\centering
\includegraphics[width = 0.999\textwidth]{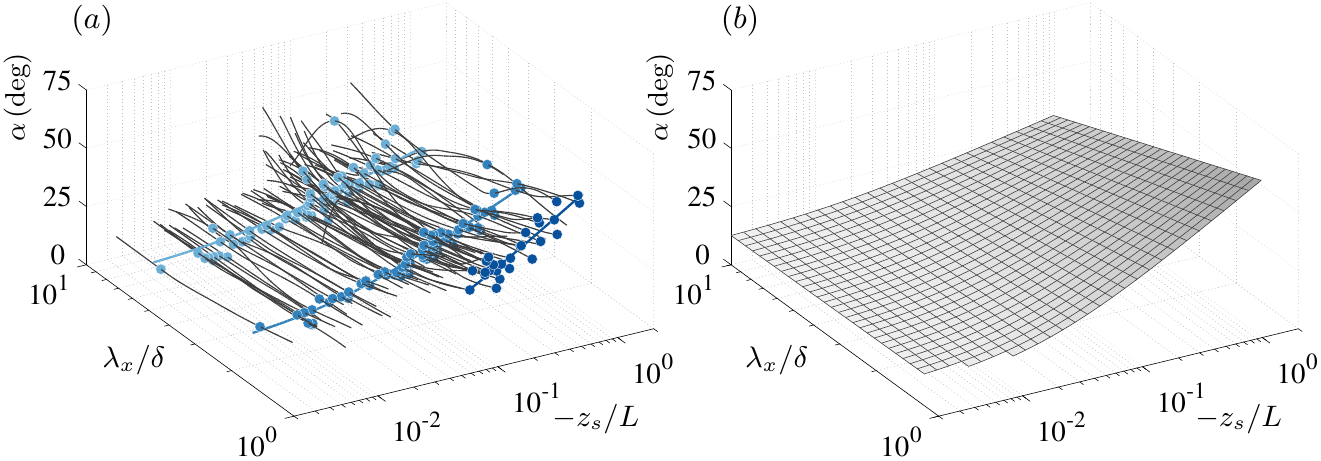}
\caption{($a$) The phase expressed as a physical inclination angle $\alpha$ as a function of wavelength $\lambda_{x}/\delta$ and stability parameter $-z_s/L$. The LCS reference height, $z_{\rm lcs}=0.90$\,m, and inclination angles are presented for distances $\Delta z = h_5 - h_1 = 4.1$\,m. The solid increasing shades of blue dots indicate the scales at $\lambda_{x}/\delta=1$; $\lambda_{x}/\delta=2$ and $\lambda_{x}/\delta=6$, which is addressed in detail in figure~\ref{fig:angle1}. The solid blue lines show the trend fitted to these data  given by the ($b$) surface fit to the data of the form (\ref{equation_zL}). }
\label{fig:phase1surf}
\end{figure}

Per the phase spectrum in figure~\ref{fig:processing}($f$), the scale-dependent phase is now analyzed for scales within the range $1 < \lambda_x/\delta < 10$ and for a similar $\Delta z$ as before: $\Delta z = h_5 - h_1 = 4.1$\,m. Larger wavelength information in the phase spectra is prone to noise issues due to the limited ensembles available for constructing the spectra, while smaller wavelengths generally have a lower coherence. Figure~\ref{fig:phase1surf}($a$) indicates the phase spectra in terms of inclination angle as a function of wavelength $\lambda_{x}/\delta$, for all different stability parameters $-z_s/L$. Note that all measured angles in figure~\ref{fig:phase1surf}($a$) are positive, corresponding to forward leaning structures. Though there are some clear outliers in these plots, certain trends are visible as evidenced by the coloured symbols which show data at constant scales of $\lambda_x/\delta$ = 1, 2 and 6. Clearly these different wavelengths exhibit different dependencies of $\alpha$ with stability parameter, with $\lambda_x/\delta = 1$ exhibiting much steeper angles $\alpha$ in the most unstable cases and markedly shallower angles as near-neutrality is approached.  In general, it is also noted that  longer structures (larger wavelength) will exhibit smaller inclination angles, especially noticeable at stronger convective conditions. Similar to \citet{chauhan2013structure} and \citet{baars2016spectral}, an extended parametric equation is fitted to the log-linear trend of convective data to model the variation of $\alpha$ with stability $z_s/L$, scale $\lambda_{x}$, following
\begin{center}
	\begin{equation}
	   		\alpha \left(\frac{z_s}{L},\frac{\lambda_x}{\delta}\right) =\alpha_{0}+C_{0}\left(\frac{\lambda_x}{\delta}\right)\ln \left(1+70\left| \frac{z_s}{L} \right| \right),
		\label{equation_zL} 
	\end{equation}
\end{center}
where $z_s = 2.5$\,m, and $\alpha_{0}$ is the constant inclination angle selected as $\alpha_{0}=12^\circ$ in the neutral surface layer, $C_0$ is a function of the outer-scaled wavelength $\lambda_x/\delta$, which will be illustrated later. A curve-fitted plane to those spectra is shown in figure~\ref{fig:phase1surf}($b$), in which this fit has the form in (\ref{equation_zL}). It should be noted from figure~\ref{fig:phase1surf}($a$) that for small near-neutral values of the stability parameter it is not always possible to compute an inclination angle $\alpha$ for the smaller wavelength ($\lambda_x/\delta = 1$) from the phase spectra since the coherence across $\Delta z$ drops below $\gamma_L^2 = 0.1$.
\begin{figure} 
\vspace{0pt}
\centering
\includegraphics[width = 0.999\textwidth]{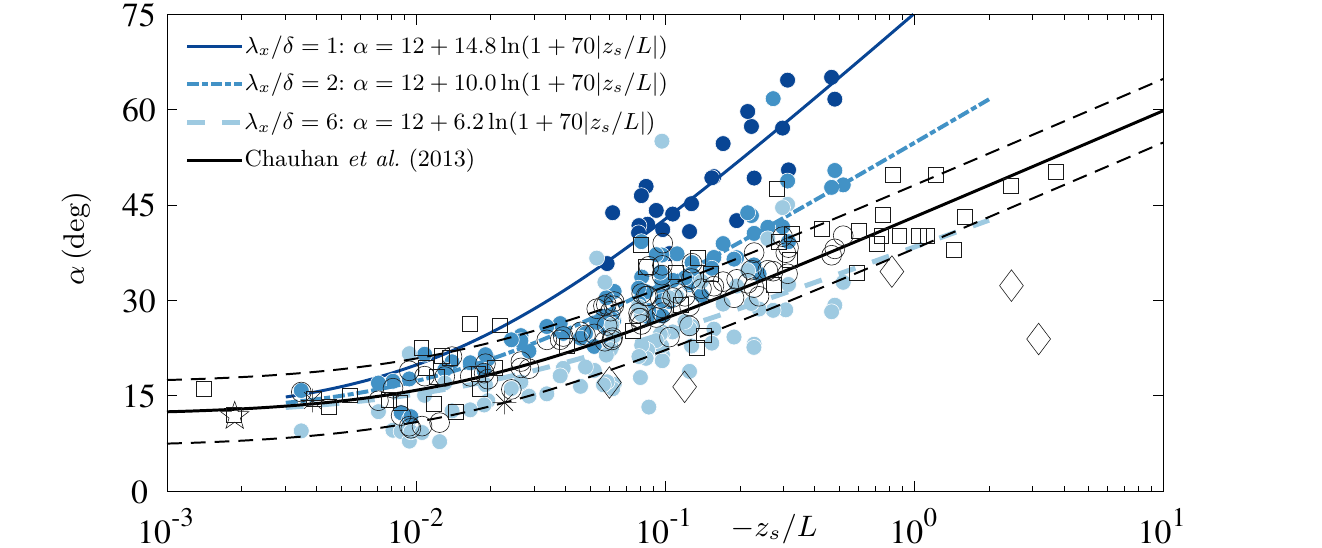}
\caption{Variation of $\alpha$ with the stability parameter $z_s/L$ ($z_s = 2.5$\,m, $z_{\rm lcs} = 0.90$\,m and $z = 5$\,m, so that $\Delta z = 4.1$\,m). The solid color lines are the fitting lines based on (\ref{equation_zL}) for the corresponding scales of $\lambda_x/\delta = 1$, 2 and 6. Current data are shown by black circles (based on a two-point correlation, including all scales). The open black square correspond to near-neutral and unstable ASL data of \citet{chauhan2013structure}; asterisks from $R_{u_{\tau}u}$ of \citet{marusic2007reynolds}; diamonds of \citet{carper2004role}; open pentagram from $R_{uu}$ of \citet{marusic2007reynolds}.}
\label{fig:angle1}
\end{figure}

Figure~\ref{fig:angle1} shows the variation of $\alpha$ as a function of the stability parameter $-z_s/L$ for three different length scales $\lambda_x/\delta = 1,2,6$. Inclination angles computed from the phase spectra are shown by the colored circles and the blue lines show the surface fit to the data given by (\ref{equation_zL}). In addition, the black open circles shown in figure~\ref{fig:angle1}, which show the inclination computed from the two-point correlations for all 11 measurement heights indicate that the current data are consistent with the similarly computed results from \citet{chauhan2013structure}, shown by the black open squares. The black solid line in figure~\ref{fig:angle1} shows the fit proposed by \citet{chauhan2013structure} based on the two-point correlation contours. As such this original fit includes \emph{all scales}, and is disproportionately skewed towards larger scale features at higher $z$. It is clear from figure~\ref{fig:angle1} that the increasing buoyancy lifts all scales to larger inclination angles, although the smaller scale structures considered ($\lambda_x/\delta = 1$) exhibit steeper angles at all values of the stability parameter as compared to the larger features ($\lambda_x / \delta = 6$). This means that the fitting parameter $C_{0}$ in (\ref{equation_zL}) increases systematically as we focus on smaller scales. The coloured curves in figure~\ref{fig:angle1} show the curve fits to the data based on the fit proposed in (\ref{equation_zL}), with the constants $\alpha_0$ and $C_0$ given in the figure legend. It should be noted here that for the limited scale range discussed here $1<\lambda_{x}/\delta<10$, all of these scales are large, and associated with the upper end of attached motions and superstructures \citep{hutchins2007evidence}. Although with increasing stability all scales are lifted compared with the neutral condition, the low end of this range ($\lambda_x/\delta = 1$) exhibits the steepest angles, reaching $\alpha \approx 70^\circ$ at $z_s/L= -1.0$. \citet{baars2016spectral} indicated that the inclination angle of the large-scale structures in the neutral laboratory boundary layer is scale independent and equal to $\alpha = 14.7^\circ$. Though the data in figure~\ref{fig:angle1} do suggest that $\alpha$ becomes scale-independent in the limit of small $|z_s/L|$, the angle seems to be closer to $\alpha \approx 12^\circ$ for the current data.

\begin{figure} 
\vspace{0pt}
\centering
\includegraphics[width = 0.999\textwidth]{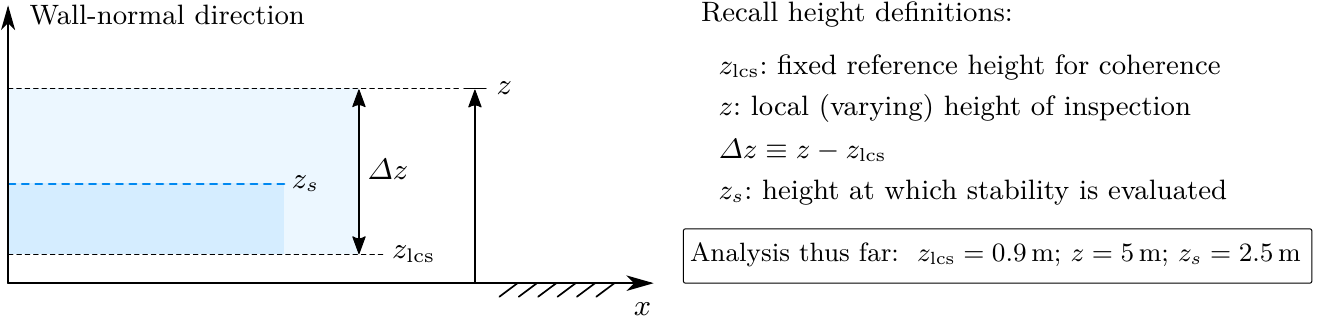}
\caption{Summary of the different heights involved in the analysis. Heights for which the analyses were performed up to this point are listed in the box.}
\label{heights}
\end{figure}
The results summarised in figure~\ref{fig:angle1} are computed for a linear coherence spectrum between the reference location $z_{\rm lcs} = 0.9$\,m and the location $z = 5$\,m ($\Delta z = 4.1$\,m). In addition, the stability parameter presented on the abscissa of figure~\ref{fig:angle1} is computed based on conditions at $z_s = 2.5$\,m. Before considering further the form of the fit described by (\ref{equation_zL}), we must consider the sensitivity to $z_s$, and $\Delta z$. Generally, the stability parameter $\vert z/L \vert$ is linearly increasing with height, which would suggest that the scale-dependent structure inclination angles $\alpha$ will also increase with wall-normal height $z$  (since as we move away from the wall, buoyancy effects will increase in dominance relative to shear). To account for this, and to form a stability parameter that better reflects the altitude at which the linear coherence is evaluated, we propose a fractional stability parameter, dubbed $z_F/L$, where $z_F$ is fixed at a constant fraction of the wall-normal offset $\Delta z$. Hence stability is always assessed at a fixed fractional height between the lower and upper probes used to compute the linear coherence spectrum. This should minimise the sensitivity of inclination angle-stability parameter relationship to changes in $\Delta z$. Since the data in figure~\ref{fig:angle1} are given based on $z_{\rm lcs} = 0.90$\,m, $z_{s} = 2.5$\,m and $z=5$\,m (figure~\ref{heights}), the fractional stability parameter $z_F$ can be assessed as
\begin{eqnarray}
    z_F &=& \overbrace{\left(\frac{z_s-z_{\rm lcs}}{z-z_{\rm lcs}}\right)}^{\rm \scalebox{0.80}{values fig.\,\ref{heights}}}\Delta z+z_{\rm lcs} = 0.39\Delta z + z_{\rm lcs}.
\label{equation_zF}
\end{eqnarray}
Maintaining $z_F$ at the value given by (\ref{equation_zF}), when changing $\Delta z$ or $z_{\rm lcs}$, ensures that we assess stability at the same fractional location 0.39$\Delta z$ when varying the height of inspection $z$. We can now substitute $z_F$ for $z_s$ in (\ref{equation_zL}),
\begin{equation}
   \alpha\left(\frac{z_F}{L},\frac{\lambda_{x}}{\delta}\right) =\alpha_{0}+C_{0}\left(\frac{\lambda_{x}}{\delta}\right) \ln\left(1+70\left|\frac{z_F}{L}\right| \right).
\label{equation_last1}
\end{equation}

\begin{figure} 
\vspace{0pt}
\centering
\includegraphics[width = 0.999\textwidth]{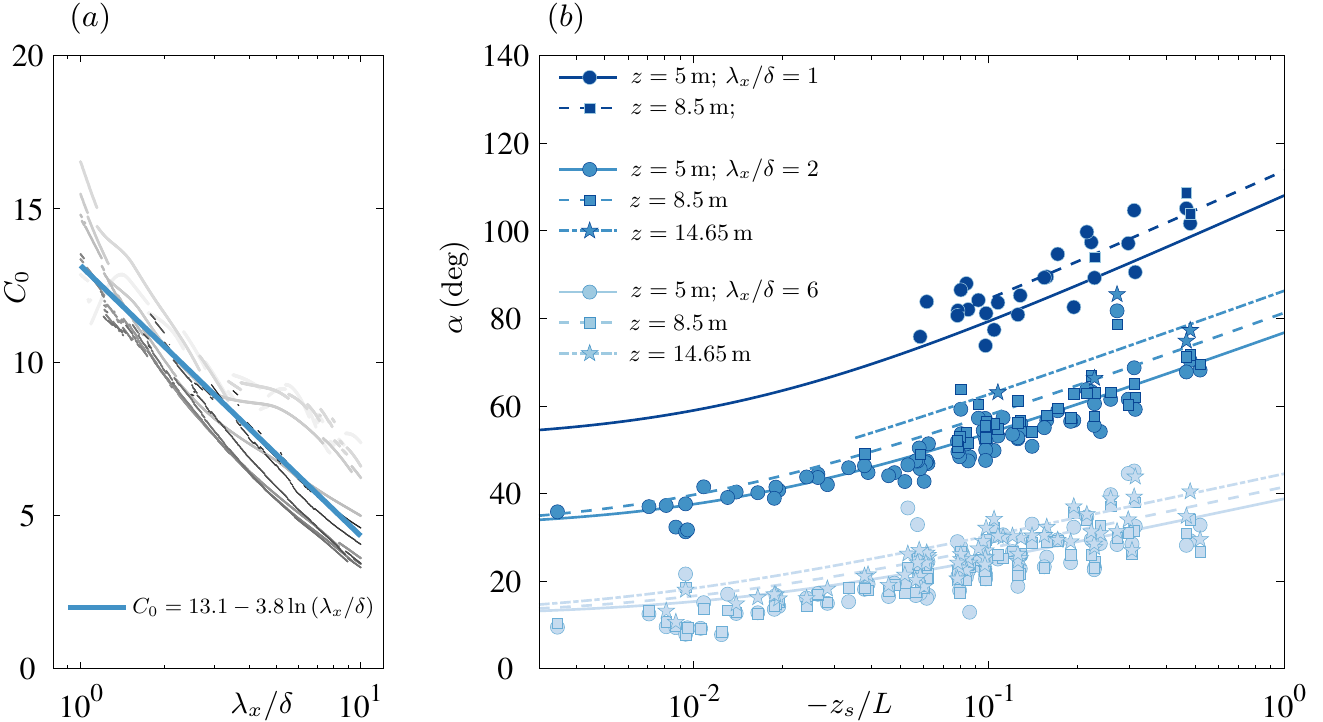}
\caption{($a$) Parameter $C_{0}$ as a function of wavelength $\lambda_x$ with increasing $\Delta z$ indicated by thinner lines and darker shades of grey. Here $z_{\rm lcs}$ remains fixed at $h_1 = 0.91$\,m and $z$ increases from 1.71\,m to 30\,m ($h_2$ to $h_{11}$), the blue solid line indicates a log-linear fit to all grey curves. ($b$) The scale-dependent $\alpha$ for $u$ as a function of stability parameter, for $\lambda_x/\delta=1,2,6$ at $h_5$, $h_7$ and $h_9$ ($z_s = 2.5$\,m and $z_{\rm lcs} = 0.90$\,m); lines come from (\ref{equation_final}) and each set of profiles is offset by $15^\circ$.}
\label{fig:c0}
\end{figure}
The grey curves in figure~\ref{fig:c0}($a$) show the fitting parameter $C_0$ from (\ref{equation_last1}) extracted from fits to the data with various $\Delta z$. Here $z_{\rm lcs}$ remains fixed at 0.90\,m and $z$ covers all heights above ranging from 1.71\,m to 30\,m, with thinner grey lines indicating lower values of $z$. We can only extract $\alpha$ from the phase spectra in situations where the coherence $\gamma_L^2$ is greater than 0.1, this condition gives rise to the sharp drops/rises in figure~\ref{fig:c0}($a$) as at different $\lambda_x/\delta$, varying subsets of data are available for the calculation of $\alpha$. In general it is noted that the fractional location for the stability height described above, does a reasonable job of collapsing the $C_0$ curves for all $\Delta z$; $C_0$ decreases with $\lambda_x/\delta$ indicating that the smaller scale structures tend to have larger inclination angles in the convective conditions. Based on the approximate collapse observed in figure~\ref{fig:c0}($a$), we can crudely approximate the $\lambda_x/\delta$ dependence of $C_0$ with,
\begin{equation}
	C_{0}=13.1-3.8\ln\left(\frac{\lambda_x}{\delta}\right),
\label{equation_last2}
\end{equation}
which is shown by the blue solid line in figure~\ref{fig:c0}($a$). Finally, by combining (\ref{equation_last2}) with (\ref{equation_last1}), we obtain the scale-dependent structural inclination angle as,
\begin{equation}
	\alpha\left(\frac{z_F}{L},\frac{\lambda_{x}}{\delta}\right) =\alpha_{0}+\left(13.1-3.8\ln\left(\frac{\lambda_{x}}{\delta}\right)\right)\ln\left(1+70\left|\frac{z_F}{L}\right|\right).
\label{equation_final}
\end{equation}

Figure~\ref{fig:c0}($b$) shows the influence of wall-normal offset $\Delta z$ (with $z_{\rm lcs}$ fixed at 0.90\,m) on the computed inclination angle $\alpha$ as a function of stability for the wavelengths $\lambda_{x}/\delta=1,2,6$ for $u$. A larger $\Delta z$ leads to higher $\alpha$, but this can be accounted for by considering the fractional stability parameter. The curves, showing (\ref{equation_final}), describe the variation of $\alpha$ with $z_s/L$, $\lambda_x/\delta$ and $\Delta z$ reasonably well. 

\begin{figure} 
\vspace{0pt}
\centering
\includegraphics[width = 0.999\textwidth]{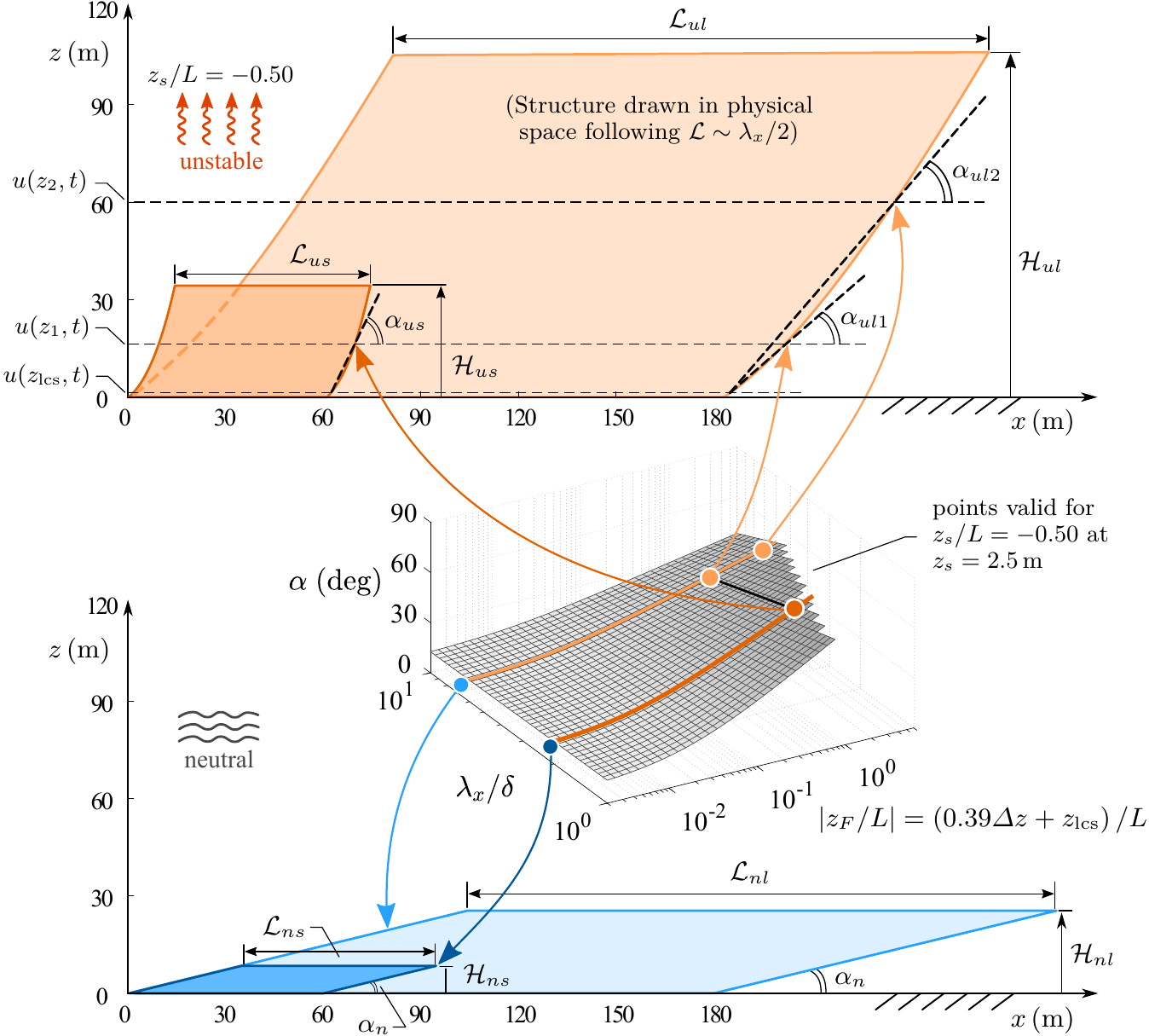}
\caption{Summary of aspect ratio $\AR^z$ and scale-dependent angle $\alpha$ in a neutral and unstable stratified ASL. Two hierarchy levels are considered, each of a different scale: $L$ and $H$ show the structure's streamwise length and wall-normal height. The subscripts' first letter $n$ or $u$ signifies a `neutral' or `unstable' condition, and the second designation $s$ or $l$ denotes `small-scale' ($\lambda_x/\delta = 2$) or `large-scale' ($\lambda_x/\delta = 6$). The dashed lines in the top figure indicate a reference location $z_{\rm lcs} = 0.90$\,m and heights $z_1 = z_{\rm lcs} + \Delta z = 15$\,m and $z_2 = z_{\rm lcs} + \Delta z = 60$\,m, for which the corresponding $\alpha$ are indicated.}
\label{figure_sketch}
\end{figure}

By way of a summary, figure~\ref{figure_sketch} shows an illustration of the scale-dependent structure inclination angle and aspect ratio for both neutral (subscript $n$) and unstable (subscript $u$) thermal stratification conditions. The illustrations show the streamwise extent of the structure (in physical space following the simplification that its length $\mathcal{L}$ scales with half the wavelength, thus $\mathcal{L} \sim \lambda_x/2$) and its wall-normal extent with height $\mathcal{H}$; the streamwise/wall-normal aspect ratio of the structure in the $x$,$z$--plane adheres to \AR$_u^z$ of figure~\ref{result_gamma}($b$). When concentrating on the neutral stability condition first (bottom figure), both scales drawn exhibit the same inclination angle $\alpha_n$. That is, the statistical inclination angle of a large-  and small-scale comprise the same forward leaning behavior.

When considering unstable thermal stratification---in the illustration solely one value is considered of $z_s/L = -0.50$ evaluated at $z_s = 2.5$\,m---the inclination angle is both scale-dependent and varies with $z$ (the procedure for plotting this structure is outlined in Appendix~A). First of all, note that the coherent wall-attached structures are still self-similar per the definition used in this paper based on the structure's streamwise wavelength relative to its wall-normal extent. But, the aspect ratio reduces compared to the neutral case and that this depends on the degree of thermal stratification per the relations shown in figure~\ref{result_gamma}($b$), thus: $(\mathcal{L}_{ns}/\mathcal{H}_{ns} = \mathcal{L}_{nl}/\mathcal{H}_{nl}) > (\mathcal{L}_{us}/\mathcal{H}_{us} = \mathcal{L}_{ul}/\mathcal{H}_{ul})$. Here subscript $s$ and $l$ refer to the small- and large-scale structure visualized, respectively. Concentrating on the inclination angle (\emph{e.g.}, the phase shift between a reference height $z_{\rm lcs}$ and a height of inspection $z$), it was found that for unstable stratification it increases with increasing $z$. Hence, $\alpha_{ul2} > \alpha_{ul1}$ and the coherent shape of the structure is characterized by a leading/trailing edge front that is curved and becomes steeper with height. Moreover, smaller scale structures exhibit steeper angles for the case of equal $\Delta z$ (\emph{e.g.}, $\alpha_{us} > \alpha_{ul1}$). The trends for this one stability condition are visualized with the `$\alpha$-surface' in the centre of figure~\ref{figure_sketch}. There the fractional stability $z_F$ is a surrogate for the $\Delta z$ trend.

\section{Conclusion}\label{sec:concl}
Wall-normal and spanwise arrays of sonic anemometers deployed in the atmospheric surface layer enable examination of the linear coherence spectrum, $\gamma_{L}^2$, as a function of the streamwise wavelength ($\lambda_{x}$), spanwise offset ($\Delta{y}$) and wall-normal offset ($\Delta{z}$). This in turn offers the opportunity to explore the three-dimensional form of the wall-attached self-similar structure for the streamwise velocity $u$, which illustrates that the self-similar wall-attached structures follow an aspect ratio of $\lambda_{x}/\Delta{z}:\lambda_{x}/\Delta{y}\approx1$ under near-neutral and unstable conditions. It is found that the aspect ratio $\lambda_x/\Delta z$ is greater for the near-neutral case, and becomes progressively smaller as instability increases. Hence similar length ($\lambda_x$) structures in unstable conditions will be taller and wider than their near-neutral counterparts. The phase of the cross-spectrum provides a scale-by-scale structure inclination angle. We find that this inclination angle is invariant with scale for the near-neutral case, but with increasing positive buoyancy  becomes increasingly scale dependent. For unstable conditions (\emph{e.g.}, $z_s/L = -0.5$), all scales are inclined at steeper angles compared to the near-neutral case, with the smaller scales with $\lambda_x/\delta = 1$ exhibiting inclination angles that are approximately twice that of larger scales of $\lambda_x/\delta = 6$.

\section*{Acknowledgements}
We acknowledge support from the National Natural Science Foundation of China (No.92052202) and the financial support from the Australian Research Council. Also, XBL was supported by a CSC scholarship (File No.\,201706180037).\\[-8pt]

\noindent \textbf{Declaration of Interests.} The authors report no conflict of interest.

\section*{Appendix A: Outline of a coherent structure}\label{sec:appA}
\begin{figure} 
\vspace{0pt}
\centering
\includegraphics[width = 0.999\textwidth]{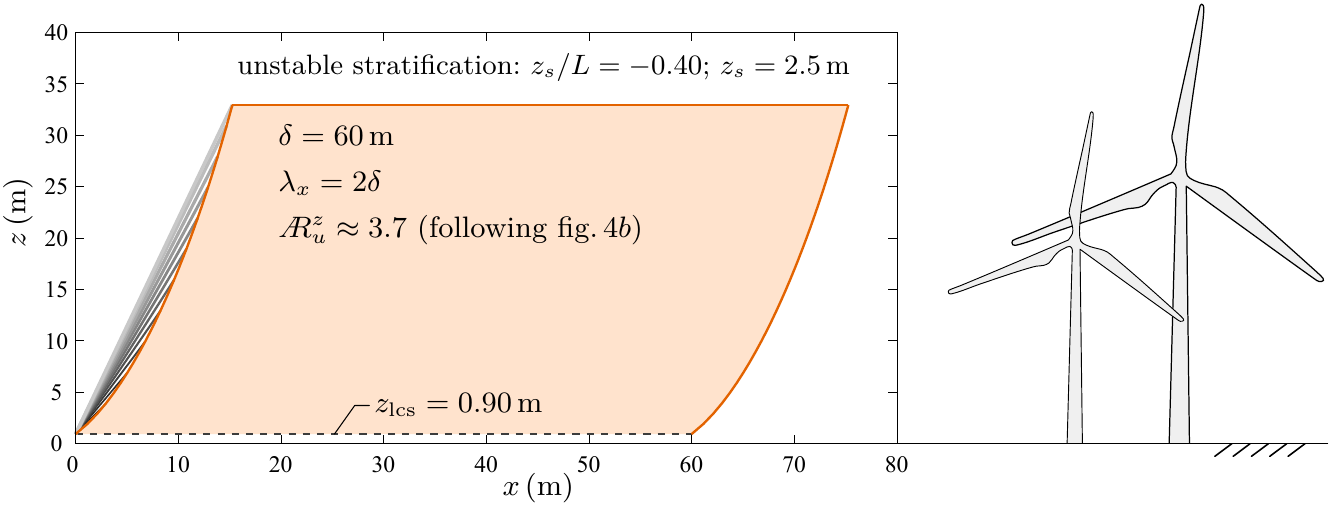}
\caption{Example of an outline of a coherent, streamwise/wall-normal structure of a particular scale and set of ASL parameters (indicated in the figure).}
\label{fig:plottingcode}
\end{figure}

Visualizing the outline of a coherent, statistical structure of a streamwise velocity fluctuation in the streamwise-wall-normal plane relies on a simple model based on (\ref{equation_final}) and (\ref{equation_zF}). The procedure for plotting a structure such as the example one in figure~\ref{fig:plottingcode} starts with the following steps:\\[-8pt]
\begin{enumerate}[labelwidth=0.65cm,labelindent=0pt,leftmargin=0.65cm,label=(\roman*),align=left]
\item \noindent A wavelength of the structure should be chosen, \emph{i.e.}, $\lambda_x = 2\delta$. Note that the structure is visualized in physical space, through the assumption that its streamwise extent spans half the wavelength, $\mathcal{L} \sim \lambda_x/2$.\\[-10pt]
\item \noindent The degree of unstable stratification should be chosen, \emph{i.e.}, $z_s/L = -0.40$ (here $z_s = 2.5$\,m).\\[-10pt]
\item \noindent A reference height, above which the structure is visualized, should be chosen.\\[-8pt]
\end{enumerate}
A structure outline is generated through considering a sequence of local heights $z$. For every height, the angle relative to the fixed reference height is determined from (\ref{equation_final}) and (\ref{equation_zF}). Note that the structure is only defined up to a height that is dictated by the aspect ratio-condition, following the trend line in figure~\ref{result_gamma}($b$). Thus, the wall-normal extent up to which the structure is defined follows from inferring the value of $\Delta z$ from \AR$_u^z \equiv \lambda_x/\Delta z = -1.9\log\left(-z_s/L\right) + 3.5$.

\bibliographystyle{jfm}
\bibliography{JFM_coherence_spectrum}

\end{document}